\documentclass[a4paper,12pt]{article}
\usepackage{graphicx,rotating,colortbl,xcolor,wrapfig,hyperref,slashed}

\hypersetup{colorlinks,bookmarksopen,bookmarksnumbered,
linkcolor=blus,pdfstartview=FitH,urlcolor=rossos,citecolor=verde}

\def\to{\rightarrow}
\def\bea{\begin{eqnarray}}
\def\eea{\end{eqnarray}}

\definecolor{Gray}{gray}{0.95}

\newcommand{\rht}{\tilde{\rho}}

\definecolor{rosso}{cmyk}{0,1,1,0.4}
\definecolor{rossos}{cmyk}{0,1,1,0.55}
\definecolor{rossoc}{cmyk}{0,1,1,0.2}
\definecolor{blu}{cmyk}{1,1,0,0.3}
\definecolor{blus}{cmyk}{1,1,0,0.6}
\definecolor{bluc}{cmyk}{1,1,0,0.1}
\definecolor{verde}{cmyk}{0.92,0,0.59,0.25}
\definecolor{verdec}{cmyk}{0.92,0,0.59,0.15}
\definecolor{verdes}{cmyk}{0.92,0,0.59,0.4}

 \def\be   {\begin{equation}}   \def\ee   {\end{equation}}
 \def\ba   {\begin{array}}      \def\ea   {\end{array}}

\font\tenrsfs=rsfs10 at 12pt
\font\sevenrsfs=rsfs7
\font\fiversfs=rsfs5
\newfam\rsfsfam
\textfont\rsfsfam=\tenrsfs
\scriptfont\rsfsfam=\sevenrsfs
\scriptscriptfont\rsfsfam=\fiversfs
\def\mathscr#1{{\fam\rsfsfam\relax#1}}

\def\circa#1{\,\raise.3ex\hbox{$#1$\kern-.75em\lower1ex\hbox{$\sim$}}\,}

\newcommand{\beq}{\begin{equation}}
\newcommand{\eeq}{\end{equation}}

 \def\kx{\kappa}
 \def\Lx{\Lambda}
 \def\lx{\lambda}

\begin{document}

\thispagestyle{empty}
\vspace{0.1cm}
\begin{center}
{\Large \bf \color{rossos} 
Multicritical points of the $O(N)$ scalar theory \\
 in $2<d<4$ for large $N$}  \\[2cm]

{\bf\large A. Katsis and N. Tetradis}  \\[5mm]

{\it Department of Physics, National and Kapodistrian University of Athens,\\ Zographou 157 84, Greece
}

\vspace{2cm}
{\large\bf\color{blus} Abstract}

\begin{quote}
We solve analytically the renormalization-group equation
for the potential of the $O(N)$-symmetric scalar theory in the large-$N$ limit 
and in dimensions $2<d<4$, in order to look for nonperturbative fixed points
that were found numerically in a recent study. We find new real solutions
with singularities in the higher derivatives of the potential at its minimum,
and complex solutions with branch cuts along the negative real axis.
\end{quote}
\end{center}

\vspace{2cm}
{{\bf Keywords}: critical phenomena, renormalization group, scalar theories, large $N$}

\newpage
\normalsize


\section{Introduction}

The $O(N)$-symmetric scalar theories have served for decades as the 
testing ground of techniques developed for the investigation 
of the critical behaviour of field theories and statistical models.
It comes, therefore, as a surprise that a recent study \cite{yabunaka} 
has found 
that their phase structure may be much more complicated that what 
had been found previously. 
In particular, it is suggested that, in 
dimensions $2<d<4$, several nonperturbative fixed points exist, which
had not been identified until now. The large-$N$ limit 
\cite{sphericallargen1,sphericallargen2,sphericallargen3,sphericallargen4,sphericallargen5,zinnjustin} offers the 
possibility to identify such fixed points analytically, without 
resorting to perturbation theory. We shall consider the theory in this limit
through the Wilsonian
approach to the renormalization group (RG) \cite{wilson}. Its various 
realizations \cite{wegner,polchinski,wetterich,chvvednic,berges} give 
consistent descriptions of the fixed-point structure of the three-dimensional
theory \cite{largenmorris}, in agreement with known results 
for the Wilson-Fisher (WF) fixed point \cite{4meps}
and the Bardeen-Moshe-Bander (BMB) endpoint of the line of 
tricritical fixed points \cite{bmb1,bmb2,bmb3}.

We shall employ the formalism of ref. \cite{wetterich}, leading to the
exact Wetterich equation for the functional RG flow of the action. For 
$N\to \infty$ the anomalous dimension of the field  
vanishes and higher-derivative terms in the action are expected to play
a minor role. This  
implies that the derivative expansion of the action \cite{derivexp1,derivexp2,derivexp3} can be truncated at the lowest order,
resulting in the local potential approximation (LPA) \cite{wegner,berges,largenmorris,aoki}. The resulting evolution equation for
the potential is exact in the sense explained in ref. \cite{largenmorris}. It 
has been analysed in refs. \cite{tet,marchais} in three dimensions.
In this work, we extend the analysis over the range $2<d<4$, in an attempt to
identify new fixed points.

\section{Evolution equation for the potential}

We consider the theory of an $N$-component scalar field $\phi^a$ with $O(N)$ symmetry in $d$ dimensions. 
We are interested in the functional RG evolution of the 
action as a function of a sharp infrared cutoff
$k$. We work within the LPA approximation, 
neglecting the anomalous dimension of the field and higher-derivative
terms in the action. We 
define $\rho=\frac{1}{2}\phi^a \phi_a $, $a=1...N$, as well as the
rescaled field $\rht=k^{2-d} \rho$. We denote
derivatives with respect to $\rht$ with primes. We focus on the
potential $U_k(\rho)$ and its dimensionless 
version $u_k(\rht)=k^{-d}U_k(\rho)$.
In the large-$N$ limit and for a sharp cutoff, the
evolution equation for the potential can be written as \cite{tet}
\be
\frac{\partial u'}{\partial t}=-2u'+(d-2)\rht \frac{\partial u'}{\partial \rht}
-\frac{NC_d}{1+u'}\frac{\partial u'}{\partial \rht},
\label{evpot} \ee
with $t=\ln(k/\Lambda)$ and $C_d^{-1}=2^d\pi^{d/2}\Gamma(d/2)$.
This equation can be considered as exact, as explained 
in ref. \cite{largenmorris}. The
crucial assumption is that, 
for $N\to \infty$, the contribution from  the radial mode 
is negligible compared to the contribution from the $N$ Goldstone modes.

The most general solution of eq. (\ref{evpot}) can be derived with 
the method of characteristics, generalizing the results of ref. 
\cite{tet}. 
It is given by the implicit relation 
\begin{eqnarray}
&& \rht
-\frac{N C_d}{d-2}
~_2F_1\left( 1,1-\frac{d}{2},2-\frac{d}{2},-u' \right)
\nonumber \\
&&~~~~
=
 e^{(2-d)t}\, G\left(u' e^{2t}\right) \,
-\frac{N C_d}{d-2} e^{(2-d)t}
~_2F_1\left( 1,1-\frac{d}{2},2-\frac{d}{2},-u' e^{2t} \right),
\label{sol} \end{eqnarray}
with $_2F_1\left( a,b,c,z \right)$ a hypergeometric function.
The function $G$ is determined by the initial condition, which 
is given by the form of the potential at the microscopic 
scale $k=\Lambda$,
i.e. $u'_\Lambda(\rht)=\Lambda^{-2} U_\Lx'(\rho)$. $G$ is determined by
inverting this relation and solving for $\rht$ in terms of $u'$, so that
$G(u')=\rht(u')|_{t=0}$. The effective action is determined by
the evolution from $k=\Lx$ to $k=0$.

We are interested in determining possible fixed points arising in the
context of the general solution (\ref{sol}). Infrared fixed points are 
approached for $k\to 0$ or $t\to -\infty$. For finite $u'$,
the last argument of the hypergeometric function in the rhs of eq. (\ref{sol})
vanishes in this limit. 
Using the expansion
\be
_2F_1\left( 1,1-\frac{d}{2},2-\frac{d}{2},-z \right)=1+\frac{d-2}{4-d}z
-\frac{d-2}{6-d}z^2+{\cal O}(z^3)
\label{expan0} \ee 
we obtain
\be
\rht
-\frac{N C_d}{d-2}
~_2F_1\left( 1,1-\frac{d}{2},2-\frac{d}{2},-u' \right)
=
 e^{(2-d)t} \left( G\left( u' e^{2t} \right) \,
-\frac{N C_d}{d-2} \right).
\label{wff} \ee
The $t$-dependence in the rhs must be eliminated for a fixed-point solution
to exist. This can be achieved for appropriate
functions $G$. For example, we may assume that the initial condition
for the potential at $k=\Lambda$ or $t=0$ is $u_\Lambda(\rht)=\lambda_\Lambda
(\rht-\kappa_\Lambda)^2/2$, so that $G(z)=\kappa_\Lambda+z/\lambda_\Lambda$.
Through the unique fine tuning $\kappa_\Lambda=N C_d/(d-2)$ the rhs vanishes
for $t\to -\infty$. The scale-independent solution, given by the
implicit relation
\be
\rht
-\frac{N C_d}{d-2}
~_2F_1\left( 1,1-\frac{d}{2},2-\frac{d}{2},-u'_* \right)
= 0,
\label{wf} \ee
describes the Wilson-Fisher fixed point. 

Near the minimum of the potential, where $u_*'\simeq 0$, we have 
\be
\rht
-\frac{N C_d}{d-2} -\frac{N C_d}{4-d}u_*'+\frac{NC_d}{6-d} (u_*')^2+{\cal O}\left((u_*')^3\right)  =0.
\label{wfexp} \ee
From this relation we can deduce that the minimum is located at 
$\rht=NC_d/(d-2)\equiv \kappa_*$, while the lowest derivatives of the potential
at this point are $u''_*(\kappa_*)=(4-d)\,(NC_d)^{-1}$, $u'''_*(\kappa_*)=2(4-d)^3/(6-d)\, (NC_d)^{-2}$.  
For large $u_*'$, we can use the expansion
\be
_2F_1\left( 1,1-\frac{d}{2},2-\frac{d}{2},-z \right)=
\Gamma\left(2-\frac{d}{2}\right)\Gamma\left(\frac{d}{2} \right)z^{\frac{d}{2}-1}
+{\cal O}\left( z^{\frac{d}{2}-2} \right)
\label{expaninf} \ee 
in order to obtain the asymptotic form of the potential:
$u_*(\rht) \sim \rht^{d/(d-2)}$.
This result is consistent with the expected critical exponent $\delta=(d+2)/(d-2)$ for vanishing anomalous dimension. 
Finally, we note that the hypergeometric function
has a pole at $z=-1$. This implies that, in the regions of negative $u'$,
the unrescaled potential $U_k\rht)$ becomes flat, with its curvature
scaling as $-k^{2}$ for $k\to 0$ \cite{convex}.

We are interested in the existence of additional fixed points. 
In $d=3$ it is known that, apart from the Wilson-Fisher
fixed point, a line of tricritical fixed points exists, 
terminating at the BMB fixed point \cite{bmb1,bmb2,bmb3}. In the following 
section we describe the flows between these fixed points in terms of the
potential, in order to obtain useful intuition for the investigation of
the case of general $d$. Our analysis extends the picture of refs.
\cite{tet,marchais} away from the fixed points. 

\section{$d=3$}

\begin{figure}
\includegraphics[width=0.7\textwidth]{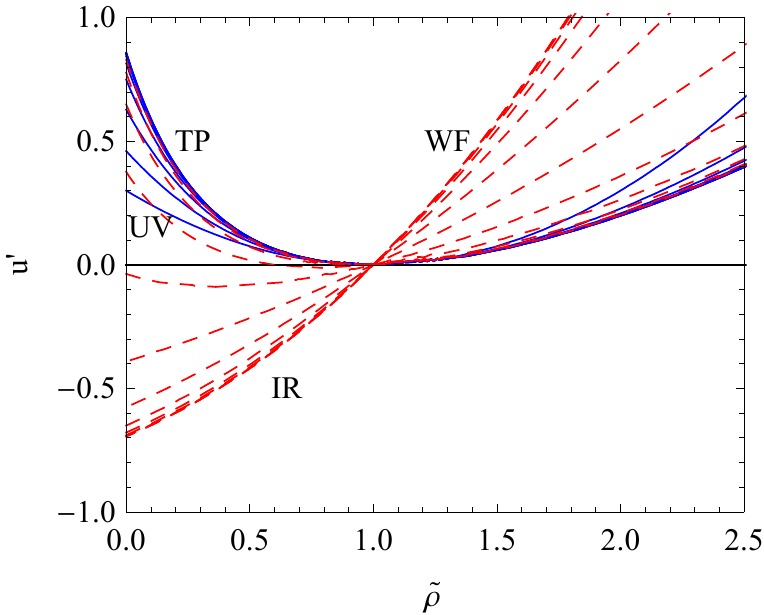}
\caption{The evolution of the potential in $d=3$ on the critical surface.
The continuous lines depict the potential during its initial
approach to the tricritical fixed point, while 
the dashed lines its subsequent
evolution towards the Wilson-Fisher fixed point.
}
\label{fig1}
\end{figure}

For $d=3$, the solution (\ref{sol}) reproduces
the one presented in ref. \cite{tet}, through use of the identity
\begin{eqnarray}
&~_2F_1\left( 1,-\frac{1}{2},\frac{1}{2},-z \right)=\sqrt{z}\,\arctan(\sqrt{z})+1
& ~~~~~z>0
\nonumber \\
&~_2F_1\left(1,-\frac{1}{2},\frac{1}{2},-z \right)=\frac{1}{2}\sqrt{-z}\ln\left(\frac{1-\sqrt{-z}}{1+\sqrt{-z}}\right)+1
& ~~~~~z<0.
\label{hyperd3} \end{eqnarray}
In order to deduce the phase diagram of the three-dimensional theory,
we consider a bare potential of the form
\be
u'_\Lx(\rht)=\lx_\Lx(\rht-\kx_\Lx )+\nu_\Lx(\rht-\kx_\Lx )^2.
\label{bare3d} \ee
The solution (\ref{sol}) can be written as
\be 
\rht-\kx_*
~_2F_1\left( 1,-\frac{1}{2},\frac{1}{2},-u' \right)
=
 e^{-t}\left[ G\left(u' e^{2t}\right) \,
-\kappa_*
~_2F_1\left( 1,-\frac{1}{2},\frac{1}{2},-u' e^{2t} \right)\right],
\label{sol3d} \ee
with $\kx_*=N/(4\pi^2)$.
The function $G(z)$ is obtained by solving eq. (\ref{bare3d}) for $\rht$ as a function of $u'_\Lx$. It is given by 
\begin{eqnarray}
G(z)&=&\kx_\Lx+\frac{1}{2\nu_\Lx} \left(- \lx_\Lx \pm
\sqrt{\lx^2_\Lx+4\nu_\Lx\, z}\right),
\end{eqnarray}
with the two branches covering different ranges of $\rht$.

Let us impose the fine tuning $\kx_\Lx=\kx_*$, which puts the theory on
the critical surface. For $\lx_\Lx\not= 0$, we have 
$G(u' e^{2t})\simeq \kx_*+u' e^{2t}/\lx_\Lx$ for $t\to -\infty$. We also
have $_2F_1\left( 1,-{1}/{2},{1}/{2},-u' e^{2t} \right)\simeq 1+u' e^{2t}$.
As a result, the
rhs of eq. (\ref{sol3d}) vanishes in this limit. The evolution leads to the 
Wilson-Fisher fixed point discussed in the previous section. 
The additional fine tuning  $\lx_\Lx= 0$ results in a different situation.
For $t\to -\infty$ the rhs of eq. (\ref{sol3d}) becomes $t$-independent and we obtain
\be 
\rht-\kx_*
~_2F_1\left( 1,-\frac{1}{2},\frac{1}{2},-u_*' \right)
=\pm \frac{1}{\sqrt{\nu_\Lx}}\sqrt{u_*'}.
\label{sol3dtr} \ee
A whole line of tricritical fixed 
points can be approached, parametrized by $\nu_\Lx$ 
\cite{marchais}. Each of them is expected to be unstable towards the 
Wilson-Fisher fixed point.

The relative stability of the fixed points can be checked explicitly by
considering the full solution (\ref{sol}). In fig. \ref{fig1} we
depict the evolution of the potential, as predicted by this expression,
for $\lx_\Lx=10^{-7}$ and $\nu_\Lx=0.3$. We have set $N C_3=1$ through
a redefinition of $\rht$ and $u'$.  
We have indicated by UV the initial form of the potential at $k=\Lx$ and
with IR its form for $k\to 0$. The continuous lines depict the potential
at various values of $t$, with step equal to $-1$, during its initial
approach to the tricritical fixed point (TP). 
The dashed lines depict its subsequent
evolution towards the Wilson-Fisher fixed point (WF).

We shall not analyse in detail the tricritical line, as this has been done 
elsewhere \cite{bmb1,bmb2,bmb3,marchais}. We note that it connects
the Gaussian fixed point, for $\nu_\Lx=0$,
with a point approached for a value of $\nu_\Lx$ for which the solution 
of eq. (\ref{sol3dtr}) diverges at the origin. This endpoint of the
tricritical line is the BMB fixed point \cite{bmb1}. The corresponding 
value of  $\nu_\Lx$ can be
derived by using the expansion (\ref{expaninf}) of the hypergeometric 
function near the origin, where the fixed-point potential diverges. 
It is given by $\nu_\Lx=2/(\pi \kx_*)$.  Taking into account our
definition of $\rho$, it can be checked that this value is consistent 
with the result of refs. \cite{bmb1,bmb2,bmb3}. 
The theory also displays first-order phase transitions if the potential
develops two minima. It was shown in ref. \cite{tet} that, for a 
bare potential of the form (\ref{bare3d}), the surface 
$\kx_\Lx=\kx_*+\lx_\Lx/\nu_\Lx$ corresponds to first-order 
phase transtions. This surface intersects the surface 
$\kx_\Lx=\kx_*$ of second-order phase transitions on the 
tricritical line $\lx_\Lx=0$.

\section{$2<d<4$}

\begin{figure}[!t]
\centering
$$
\includegraphics[width=0.5\textwidth]{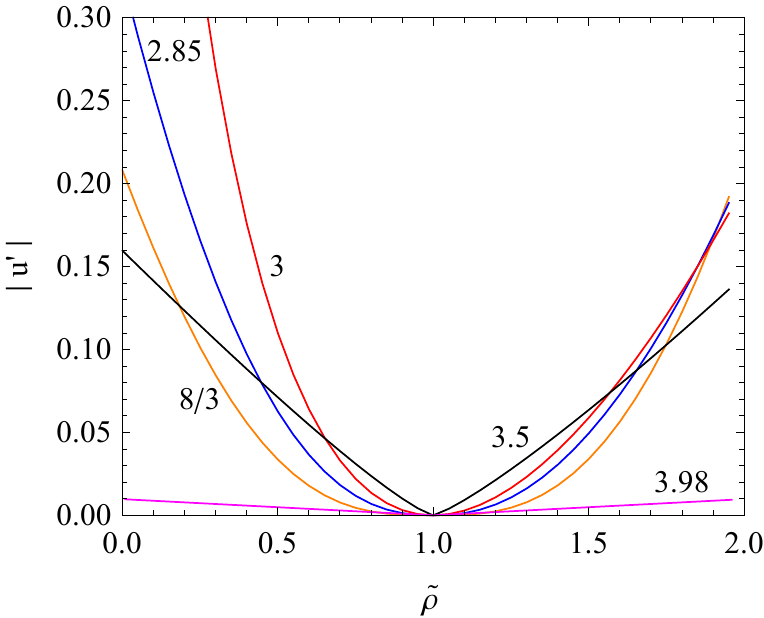}\qquad 
\includegraphics[width=0.5\textwidth]{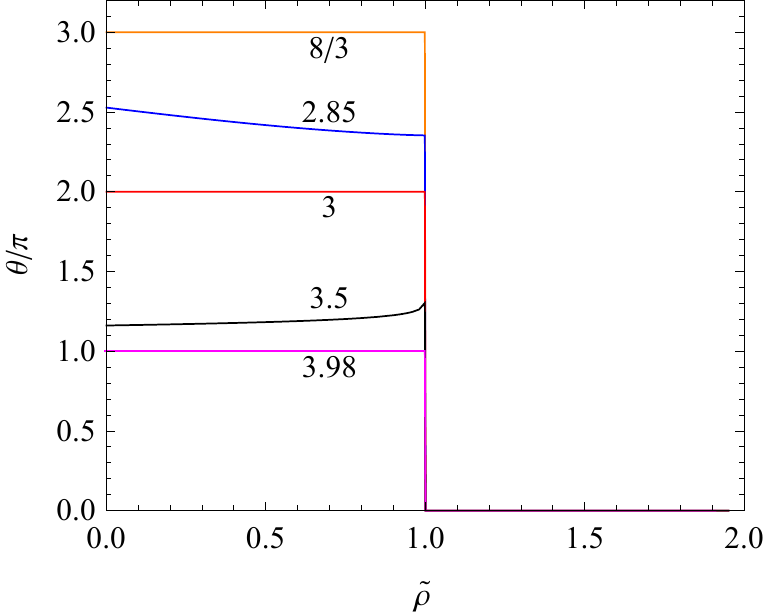}\qquad $$
\caption{\em 
The absolute value and the argument of the complex potential at the multicritical point, 
in dimensions $d=3.98,3.5,3,2.85,8/3$.}
\label{fig2}
\end{figure}

We next turn to the search for new infrared fixed points with more than one
relevant directions.
The presence of the Gaussian fixed point, with $u'=0$, 
is obvious from eq. (\ref{evpot}). For any nontrivial fixed point, the
rhs of eq. (\ref{sol}) must become independent of $t$ 
in the limit $t\to -\infty$. For $d>2$ the hypergeometric function
can be approximated through the
asymptotic expansion (\ref{expan0}) in this limit. 
An expression independent of $t$ requires
an appropriate choice of the function $G$, determined through the initial condition.
More precisely, we must have 
$G(z)- \kappa_* \sim z^{\frac{d-2}{2}}$
 for $z \to 0$. This can be achieved through
an initial condition of the form
\be u_\Lambda(\rht) 
=\frac{d-2}{d}\nu_\Lambda
(\rht-\kappa_\Lambda)^{\frac{d}{d-2}},
\label{initcondi} \ee
where the 
parametrization of the multiplicative constant has been introduced for 
later convenience. We obtain 
\be
G(z)=\kappa_\Lambda+\nu_\Lambda^{\frac{2-d}{2}} z^{\frac{d-2}{2}}.
\label{gzz} \ee
The tuning 
$\kappa_\Lambda= \kappa_*$
 results in a fixed-point potential given by the solution of
\be
\rht
-\frac{N C_d}{d-2}
~_2F_1\left( 1,1-\frac{d}{2},2-\frac{d}{2},-u'_* \right)
=\nu_\Lambda^{\frac{2-d}{2}} (u'_*)^{\frac{d-2}{2}},
\label{multi} \ee
where we have taken the limit $t\to -\infty$ with finite $u'_*$. 
The fixed-point potential has $u'_*=0$ at $\rht=\kappa_*$, similarly to the
bare potential $u_\Lambda$. 

It is apparent from eqs. (\ref{initcondi}), (\ref{multi}) 
that a nonsingular real potential for
all values of $\rht\geq 0$ can be obtained only if $d/(d-2)$ takes positive 
integer values $n$, i.e. at dimensions $d=2n/(n-1)$. If we require $2<d<4$, we 
have $n \geq 3$.
Approaching a fixed point requires, apart from the tuning
of $\kappa_\Lambda$, the absence of all 
terms $(\rht-\kappa_\Lambda)^m$ with $1<m<n$ in the bare potential. 
(The absence of the term with $m=1$ is equivalent to the tuning of $\kappa_\Lambda$.) This means that the fixed point at a given dimension
$d=2n/(n-1)$ has $n-1$ relevant directions and can be characterized as
a multicritical point. For $\rht < \kappa_*$ the form of the fixed-point potential 
depends on $n$. For $n$ odd we have $u'_*(\rht)>0$ for 
$\rht<\kappa_*$, while for $n$ even we have $u'_*(\rht)<0$. In the second
case, the potential at the origin is constrained by the pole in 
the hypergeometric function to satisfy $u'_*(0)>-1$. 

For $d\not= 2n/(n-1)$, the initial condition (\ref{initcondi}) and the solution  
(\ref{multi}) develop certain pathologies. For $\rht > \kx_*$, the potential
is real. For $\rht\gg \kappa_*$ we have 
$u'_*\gg 1$, so that the hypergeometric
function in eq. (\ref{multi}) has the expansion (\ref{expaninf}).
As we discussed above, we obtain 
the asymptotic form of the potential
$u_*(\rht) \sim \rht^{d/(d-2)}$ and a critical exponent $\delta=(d+2)/(d-2)$. 
However, divergencies in the higher derivatives of both the bare and fixed-point potentials appear as one approaches the point
$\rht=\kx_*$, at which $u'_\Lx=u'_*=0$. 
The situation is more problematic for $\rht<\kx_*$, where eqs. (\ref{initcondi}), (\ref{multi}) indicate that the potential must become complex.
This leads to the conclusion that  
a continuous range of real fixed-point solutions as a function of $d$ 
does not exist in the large-$N$ limit. 

It must be pointed out that a real solution can be constructed through an
initial condition of the form
\be u_\Lambda(\rht) 
=\pm \frac{d-2}{d}\nu_\Lambda
|\rht-\kappa_\Lambda|^{\frac{d}{d-2}},
\label{initcondib} \ee
where the positive sign is used for $\rht>\kx_*$, while the choice 
is ambiguous for $\rht<\kx_*$. Both signs lead to real potentials, but
for both choices the potentials
are nonanalytic at $\rht=\kx_*$. It cannot be excluded that the
nonanalyticity has a physical origin. On the other hand, 
it is not possible to have a continuous dependence of the fixed-point
potentials on $d$. The real and continuous solutions at 
$d = 2n/(n-1)$ result from initial conditions given by (\ref{initcondib})
with one of the two signs, but
switch from one sign to the other as $n$ is increased.

The only way to preserve a notion of analyticity and a continuous 
dependence on $d$ seems to be to consider a continuation of the 
potential in the complex plane. Even though we cannot offer a physical
interpretation of the potential, such a construction is interesting because
it may be linked to the picture presented in ref. \cite{yabunaka}. 
There, it is found that fixed-point solutions that exist for a continuous 
range of increasing values of $N$ collide with each other at
some critical value $N_c(d)$ and disappear, consistently with what has been seen 
through
the $\epsilon$-expansion \cite{stergiou}. The collision of two-fixed points is
expected to cause them to move into the complex plane \cite{kaplan}. In this
sense, 
the presence of complex fixed-point solutions for the full potential at
$N\to \infty$ would be consistent with the findings of ref. \cite{yabunaka}. 

In fig. \ref{fig2} we present complex solutions of the fixed-point equation
(\ref{multi}) for $\nu_\Lx=0.3$. We have set $N C_d=1$ through a 
redefinition of $\rht$ and $u'$. 
The left plot depicts the absolute value and the right plot the 
argument of the complex potential at the multicritical point.  
For $\rht>\kx_*=1$ the solution is real and the argument vanishes.
For $\rht<1$ the solution is real and continuous only at $d=3$ and $8/3$, 
and in general at $d=2n/(n-1)$, as discussed earlier. For any other 
value of $d$, the bare and fixed-point potentials have branch cuts
along the negative real axis. In fig. \ref{fig2} we depict the 
argument of the potential as the negative real axis
is approached from above. The argument has the opposite sign 
when the negative real axis is approached from below.
The potential is discontinuous as the
negative real axis is crossed, apart from at
$d=2n/(n-1)$. On the other hand, there is a continuity in the dependence 
of the potential on $d$. In particular, for
$\rht < \kx_\Lx$, the potential switches
automatically from solutions with $u'>0$ to ones with $u'<0$ and back,
as $n$ is increased.

\section{Conclusions}

Our analysis aimed at examining the presence of nonperturbative 
fixed-point solutions of the $O(N)$-symmetric scalar theory in dimensions
$2<d<4$ for $N\to \infty$. The motivation arose through the findings
of ref. \cite{yabunaka}, which indicate the presence of previously
unknown fixed-point solutions for finite $N$. Some of the new solutions 
collide with each other at some critical value $N_c(d)$ and disappear. 
One expects the presence of complex solutions beyond this critical value 
\cite{kaplan}. However, some novel real solutions are expected to persist in the limit
$N\to \infty$ \cite{priv}. Our aim was to identify them through an
analytical treatment of the RG equation. In this respect, a crucial point is 
our assumption about what constitutes the leading contribution for large $N$. 
For vanishing anomalous dimension, and under the assumption that higher-derivative 
terms in the action can be neglected, the exact Wetterich equation is
reduced to a partial differential equation for the potential \cite{wetterich}. 
The renormalization of the potential is induced by a term proportional to $N$,
arising from the contributions of the Goldstone modes, and a term arising from 
the contribution of the unique radial mode. Our large-$N$ approximation 
consists in neglecting the second term. 

We presented the exact solution (\ref{sol}) of the large-$N$ equation 
(\ref{evpot})
for the evolution of the
potential towards the infrared, starting from an initial condition at
an ultraviolet energy scale. The presence of critical points in dimensions
$2<d<4$, the
necessary fine tunings of the initial condition in order to approach them
during the evolution, as well as their relative stability, can be deduced 
from  eq. (\ref{sol}) by specifying the function $G(z)$.  
Our analysis of the previous two sections reproduced the known critical and
multicritical points, including the Wilson-Fisher fixed point and the
BMB fixed point. However, it did not reveal any new analytic solutions. 
Even though we used a sharp cutoff for our analysis, we expect similar
results for other cutoff functions. 
For example, the 
three-dimensional fixed-point structure that we identified is the same as the
one found in ref. \cite{marchais} with a different cutoff.

By continuing the potential in the complex plane, we obtained a
class of solutions with a branch-cut discontinuity along the negative
real axis and a continuous dependence on $d$. These solutions become
real at specific values of $d$, thus reproducing the known multicritical
points. The presence of complex fixed points is consistent with the 
finding of ref. \cite{yabunaka} that
fixed-point solutions that exist for finite 
$N$ collide with each other at some critical value $N_c(d)$ and disappear. 
On the other, it is expected that some of the real solutions presented in
ref. \cite{yabunaka} survive for $N\to \infty$ \cite{priv}. No such solutions
were found through our analysis. The only new real solutions we found 
display discontinuities or singularities in the higher derivatives of the potential
at its minimum. They can be obtained from an initial condition given
by eq. (\ref{initcondib}), for both signs, as discussed in the 
previous section. A natural question is
whether some of the numerical solutions presented in ref. \cite{yabunaka}
display similar discontinuities or singularities, so that they can 
be identified with our solutions. Another possibility is that our 
assumption that the radial mode gives a contribution subleading in $1/N$ is
violated by the novel solutions \cite{yabunaka,priv}.

\subsubsection*{Acknowledgments}
N.T. would like to thank B. Delamotte, M. Moshe, A. Stergiou, S. Yabunaka
for useful discussions. A big part of this work was carried out while N.T. was 
visiting the Theoretical Physics Department of CERN.

    \end{document}